\documentclass[fleqn,twoside]{article}                                                 
\usepackage{espcrc2}                                                                   
\usepackage{graphicx}                                                                  
                                                          
\newcommand{\AmS}{{\protect\the\textfont2                                              
  A\kern-.1667em\lower.5ex\hbox{M}\kern-.125emS}}                                      
\newcommand{\ud}{\mathrm{d}}

\title{Classical and Quantum Ensembles via Multiresolution.\\
I. BBGKY Hierarchy 
}
\author{Antonina~N.~Fedorova and Michael~G.~Zeitlin
\address{IPME RAS, St.~Petersburg, 
V.O. Bolshoj pr., 61, 199178, Russia\\
E-mail: zeitlin@math.ipme.ru, anton@math.ipme.ru\\
http://www.ipme.ru/zeitlin.html,
http://www.ipme.nw.ru/zeitlin.html}}

\begin{document}
%%%%%%%%%%%%%%%%%%%%%%%%%%%%%%%%%%%%%%%%%%%%%%%%%%%%%%%%%%%%%%%%%%%%%%%%%%%
\thispagestyle{empty}

\begin{center}
\begin{tabular}{p{130mm}}

\begin{center}
{\bf\Large CLASSICAL AND QUANTUM ENSEMBLES}\\
\vspace{5mm}

{\bf\Large VIA MULTIRESOLUTION.} \\
\vspace{5mm}

{\bf\Large I. BBGKY HIERARCHY}\\

\vspace{1cm}

{\bf\Large Antonina N. Fedorova, Michael G. Zeitlin}\\

\vspace{1cm}

{\bf\it
IPME RAS, St.~Petersburg,
V.O. Bolshoj pr., 61, 199178, Russia}\\
{\bf\large\it e-mail: zeitlin@math.ipme.ru}\\
{\bf\large\it e-mail: anton@math.ipme.ru}\\
{\bf\large\it http://www.ipme.ru/zeitlin.html}\\
{\bf\large\it http://www.ipme.nw.ru/zeitlin.html}
\end{center}

\vspace{1cm}
\begin{center}
\begin{tabular}{p{100mm}}
A fast and efficient numerical-analytical approach
is proposed for modeling complex behaviour
in the BBGKY hierarchy  of kinetic equations.
We construct the multiscale representation for hierarchy of reduced distribution functions
in the variational approach and multiresolution 
decomposition in polynomial tensor algebras of high-localized states.
Numerical modeling shows the creation of
various internal
structures from localized modes, which are related to localized or chaotic type
of behaviour and
the corresponding patterns (waveletons) formation. 
The localized pattern is a model for energy confinement state (fusion) in plasma.
\end{tabular}
\end{center}

\vspace{20mm}

\begin{center}
{\large Presented at IX International Workshop on Advanced} \\
{\large Computing and Analysis Techniques in Physics Research}\\
{\large ACAT03, December, 2003, KEK, Tsukuba, Japan}

\vspace{5mm}

{\large Nuclear Instruments and Methods in Physics Research A, in press}
\end{center}
\end{tabular}
\end{center}
\newpage

%%%%%%%%%%%%%%%%%%%%%%%%%%%%%%%%%%%%%%%%%%%%%%%%%%%%%%%%%%%%%%%%%%%%%%%%%%%
\begin{abstract} 
A fast and efficient numerical-analytical approach
is proposed for modeling complex behaviour
in the BBGKY hierarchy  of kinetic equations.
We construct the multiscale representation for hierarchy of reduced distribution functions
in the variational approach and multiresolution 
decomposition in polynomial tensor algebras of high-localized states.
Numerical modeling shows the creation of
various internal
structures from localized modes, which are related to localized or chaotic type
of behaviour and
the corresponding patterns (waveletons) formation. 
The localized pattern is a model for energy confinement state (fusion) in plasma.
\vspace{1pc}
\end{abstract}

\maketitle

\section{INTRODUCTION}

Kinetic theory is an important part of general statistical physics
related to phenomena
which cannot be understood on the thermodynamic or fluid models level [1].
In these two papers we consider the applications of a new nu\-me\-ri\-cal/analytical 
technique based on wavelet analysis approach for 
calculations related to the description of complex (non-equilibrium) behaviour  
of the corresponding classical and quantum ensembles.
The classical ensembles in this part are considered in the framework 
of the general BBGKY hierarchy and the quantum ones in part 2 in the Wigner-Weyl approach [1]. 
We restrict ourselves to the rational/polynomial type of
nonlinearities (with respect to the set of all dynamical variables)
that allows  to use our results from 
[2], which are based on the so called multiresolution
framework [3] and the variational formulation of initial nonlinear (pseudodifferential) problems.
Wavelet analysis is a set of mathematical
methods which give  a possibility to work with well-localized bases in
functional spaces and provide the maximum sparse forms  for the general 
type of operators (differential,
integral, pseudodifferential) in such bases. 
It provides the best possible rates of convergence and minimal complexity 
of algorithms inside and, 
as a result, saves CPU time and HDD space [3].
Our main goals are an attempt of classification and construction 
of a possible zoo of nontrivial (meta) stable states:
(a) high-localized (nonlinear) eigenmodes, (b) complex (chaotic-like or entangled) 
patterns, (c) localized (stable) patterns (waveletons). 
In case (c) an energy is distributed during some time (sufficiently large) 
between only a few  localized modes (from point (a)). 
We believe, it is a good image for plasma in a fusion state (energy confinement).
Our construction of cut-off of the infinite system of equations 
is based on some criterion of convergence of the full solution.
This criterion
is based on a natural norm in the proper functional space, 
which takes into account (non-perturbatively) the underlying 
multiscale structure of complex statistical dynamics.
In Sec. 2  the kinetic BBGKY hierarchy is formulated.
In Sec. 3 we present the explicit analytical construction of solutions of
the hierarchy, which is based on tensor algebra
extensions of bases generated by the hidden 
multiresolution
structure and proper variational formulation leading 
to an algebraic parametrization of the solutions.
So, our approach resembles Bogolyubov's and related approaches 
but we don't use any perturbation technique (like virial expansion)
or linearization procedures.
Numerical modeling as in general case as in particular cases of the  Vlasov-like equations 
shows the creation of
various internal
structures from localized bases modes, which demonstrate 
the possiblity of existence  of  
(metastable) pattern formation.

\section{BBGKY HIERARCHY}

Let $M$ be the phase space of an ensemble of $N$ particles ($ {\rm dim}M=6N$)
with coordinates
$x_i=(q_i,p_i), \quad i=1,...,N, \quad
q_i=(q^1_i,q^2_i,q^3_i)\in R^3,\quad
p_i=(p^1_i,p^2_i,p^3_i)\in R^3,\quad
q=(q_1,\dots,q_N)\in R^{3N}$.
Individual and collective measures are: 
$
\mu_i=\ud x_i=\ud q_i\ud p_i,\quad \mu=\prod^N_{i=1}\mu_i
$.
Our constructions can be applied to the following general Hamiltonians:
$$
H_N=
\sum^N_{i=1}\Big(\frac{p^2_i}{2m}+U_i(q)\Big)+
\sum_{1\leq i\leq j\leq N}U_{ij}(q_i,q_j)  
$$
where the potentials 
$U_i(q)=U_i(q_1,\dots,q_N)$ and $U_{ij}(q_i,q_j)$
are restricted to rational functions of the coordinates.
Let $L_s$ and $L_{ij}$ be the Liouvillean operators (vector fields)
\begin{eqnarray}
L_s=\sum^s_{j=1}\Big(\frac{p_j}{m}\frac{\partial}{\partial q_j}-
\frac{\partial U_j}{\partial q}\frac{\partial}{\partial p_j}\Big)-
\sum_{1\leq i\leq j\leq s}L_{ij}
\end{eqnarray}
\begin{eqnarray}
L_{ij}=\frac{\partial U_{ij}}{\partial q_i}\frac{\partial}{\partial p_i}+
\frac{\partial U_{ij}}{\partial q_j}\frac{\partial}{\partial p_j}\nonumber
\end{eqnarray}
and
$
F_N(x_1,\dots,x_N;t)
$
be the hierarchy of $N$-particle distribution function.
satisfying the standard BBGKY--hierarchy ($V$  is the volume) [1]:
\begin{eqnarray}
\frac{\partial F_s}{\partial t}+L_sF_s=
\frac{1}{V^s}\int\ud\mu_{s+1}
\sum^s_{i=1}L_{i,s+1}F_{s+1}
\end{eqnarray}
In most cases, one is interested in a representation of the form 
$
F_k(x_1,\dots,x_k;t)=\prod^k_{i=1}F_1(x_i;t)+G_k(x_1,\dots,x_k;t)
$
where $G_k$ are correlators. Additional reductions often lead to 
simplifications, the simplest one, $G_k=0$, corresponding to the 
Vlasov approximation.
Such physically motivated ansatzes for $F_k$
formally replace the linear (in $F_k$) and pseudodifferential (in general case)
infinite system (2) by
a finite-dimensional but nonlinear system with
polynomial  nonlinearities (more exactly, multilinearities [3]).
Our key point in the following consideration is the proper nonperturbative 
generalization of the perturbative multiscale approach of Bogolyubov.

\section{MULTISCALE ANALYSIS}

The infinite hierarchy of distribution functions satisfying system (2)
in the thermodynamical limit is:
\begin{equation}
F=\{F_0,F_1(x_1;t),\dots,
F_N(x_1,\dots,x_N;t),\dots\}
\end{equation}
where
$F_p(x_1,\dots, x_p;t)\in H^p$,
$H^0=R,\quad H^p=L^2(R^{6p})$ (or any different proper functional spa\-ce), $F\in$
$H^\infty=H^0\oplus H^1\oplus\dots\oplus H^p\oplus\dots$
with the natural Fock space like norm 
(guaranteeing the positivity of the full measure):
\begin{eqnarray}
(F,F)=F^2_0+\sum_{i}\int F^2_i(x_1,\dots,x_i;t)\prod^i_{\ell=1}\mu_\ell.
\end{eqnarray}
First of all we consider $F=F(t)$ as a function of time only,
$F\in L^2(R)$, via
multiresolution decomposition which naturally and efficiently introduces 
the infinite sequence of the underlying hidden scales [3].
Because the affine
group  of translations and dilations 
generates multiresolution approach, this
method resembles the action of a microscope. We have the contribution to
the final result from each scale of resolution from the whole
infinite scale of spaces. 
We consider a multiresolution decomposition of $L^2(R)$ [3]
(of course, we may consider any different and proper for some particular case functional space)
which is a sequence of increasing closed subspaces $V_j\in L^2(R)$ 
(subspaces for 
modes with fixed dilation value):
\begin{equation}
...V_{-2}\subset V_{-1}\subset V_0\subset V_{1}\subset V_{2}\subset ...
\end{equation}
The closed subspace
$V_j (j\in {\bf Z})$ corresponds to  the level $j$ of resolution, 
or to the scale j
and satisfies
the following properties:
let $W_j$ be the orthonormal complement of $V_j$ with respect to $V_{j+1}$: 
$
V_{j+1}=V_j\bigoplus W_j.
$
Then we have the following decomposition:
\begin{eqnarray}
\{F(t)\}=\bigoplus_{-\infty<j<\infty} W_j =
\overline{V_c\displaystyle\bigoplus^\infty_{j=0} W_j}
\end{eqnarray}
in case when $V_c$ is the coarsest scale of resolution.
The subgroup of translations generates a basis for the fixed scale number:
$
{\rm span}_{k\in Z}\{2^{j/2}\Psi(2^jt-k)\}=W_j.
$
The whole basis is generated by action of the full affine group:
\begin{eqnarray}
&&{\rm span}_{k\in Z, j\in Z}\{2^{j/2}\Psi(2^jt-k)\}=\\
&&{\rm span}_{k,j\in Z}\{\Psi_{j,k}\}
=\{F(t)\}\nonumber
\end{eqnarray}
Let the sequence $\{V_j^t\}, V_j^t\subset L^2(R)$ 
corresp\-ond to multiresolution analysis on the time ax\-is, 
$\{V_j^{x_i}\}$ correspond to multiresolution analysis for coordinate $x_i$,
then
$
V_j^{n+1}=V^{x_1}_j\otimes\dots\otimes V^{x_n}_j\otimes  V^t_j
$
corresponds to the multiresolution analysis for 
the $n$-particle distribution function 
$F_n(x_1,\dots,x_n;t)$.
E.g., for $n=2$: $V^2_0=\{f:f(x_1,x_2)=$
$
\sum_{k_1,k_2}a_{k_1,k_2}\phi^2(x_1-k_1,x_2-k_2),\ 
a_{k_1,k_2}\in\ell^2(Z^2)\},
$
where 
$
\phi^2(x_1,x_2)=\phi^1(x_1)\phi^2(x_2)=\phi^1\otimes\phi^2(x_1,x_2),
$
and $\phi^i(x_i)\equiv\phi(x_i)$ form a multiresolution basis corresponding to
$\{V_j^{x_i}\}$.
If $\{\phi^1(x_1-\ell)\},\ \ell\in Z$ form an orthonormal set, then 
$\phi^2(x_1-k_1, x_2-k_2)$ form an orthonormal basis for $V^2_0$.
So, the action of the affine group generates multiresolution representation of
$L^2(R^2)$. After introducing the detail spaces $W^2_j$, we have, e.g. 
$
V^2_1=V^2_0\oplus W^2_0.
$
Then the 
3-component basis for $W^2_0$ is generated by 
the translations of three functions:
$
\Psi^2_1=\phi^1(x_1)\otimes\Psi^2(x_2), \Psi^2_2=\Psi^1(x_1)\otimes\phi^2(x_2),
\Psi^2_3=\Psi^1(x_1)\otimes\Psi^2(x_2).
$
Also, we may use the rectangle lattice of scales and one-dimensional wavelet
decomposition:
$$
f(x_1,x_2)=\sum_{i,\ell;j,k}\langle f,\Psi_{i,\ell}\otimes\Psi_{j,k}\rangle
\Psi_{j,\ell}\otimes\Psi_{j,k}(x_1,x_2)
$$
where the basis functions $\Psi_{i,\ell}\otimes\Psi_{j,k}$ depend on
two scales $2^{-i}$ and $2^{-j}$.
We obtain our multiscale\-/mul\-ti\-re\-so\-lu\-ti\-on 
representations (formulae (11) below) 
via the variational wavelet approach for 
the following formal representation of the BBGKY system (9) 
(or its finite-dimensional nonlinear approximation for the 
$n$-particle distribution functions) 
with the
corresponding obvious constraints on 
the distribution functions.
Let $L$ be an arbitrary (non)li\-ne\-ar dif\-fe\-ren\-ti\-al\-/\-in\-teg\-ral operator 
 with matrix dimension $d$
(finite or infinite), 
which acts on some set of functions
from $L^2(\Omega^{\otimes^n})$:  
$\quad\Psi\equiv\Psi(t,x_1,x_2,\dots)=\Big(\Psi^1(t,x_1,x_2,\dots), \dots$,
$\Psi^d(t,x_1,x_2,\dots)\Big)$,
 $\quad x_i\in\Omega\subset{\bf R}^6$, $n$ is the number of particles:
{\setlength\arraycolsep{1pt}
\begin{eqnarray}
&&L\Psi\equiv L(Q,t,x_i)\Psi(t,x_i)=0,\\
&&Q\equiv Q_{d_0,d_1,d_2,\dots}(t,x_1,x_2,\dots,\nonumber\\
&&\partial /\partial t,\partial /\partial x_1,
\partial /\partial x_2,\dots,\int \mu_k)=\nonumber\\
&&\sum_{i_0,i_1,i_2,\dots=1}^{d_0,d_1,d_2,\dots}
q_{i_0i_1i_2\dots}(t,x_1,x_2,\dots)\nonumber\\
&&\Big(\frac{\partial}{\partial t}\Big)^{i_0}\Big(\frac{\partial}{\partial x_1}\Big)^{i_1}
\Big(\frac{\partial}{\partial x_2}\Big)^{i_2}\dots\int\mu_k\nonumber 
\end{eqnarray}
}
Let us consider now the $N$ mode approximation for the solution as 
the following ansatz:
\begin{eqnarray}
\Psi^N(t,x_1,x_2,\dots)=
\sum^N_{i_0,i_1,i_2,\dots=1}a_{i_0i_1i_2\dots}\\
 A_{i_0}\otimes 
B_{i_1}\otimes C_{i_2}\dots(t,x_1,x_2,\dots)\nonumber
\end{eqnarray}
We shall determine the expansion coefficients from the following conditions
(different related variational approaches are considered in [2]):
\begin{eqnarray}
&&\ell^N_{k_0,k_1,k_2,\dots}\equiv 
\end{eqnarray}
$$
\int(L\Psi^N)A_{k_0}(t)B_{k_1}(x_1)C_{k_2}(x_2)\ud t\ud x_1\ud x_2\dots=0
$$
Thus, we have exactly $dN^n$ algebraical equations for  $dN^n$ unknowns 
$a_{i_0,i_1,\dots}$.
So, the solution is parametrized by the solutions of two sets of 
reduced algebraical
problems, one is linear or nonlinear
(depending on the structure of the operator $L$) and the rest are linear
problems related to the computation of the coefficients of the algebraic equations (10).
which can be found 
by using the
compactly supported wavelet basis functions for the expansions (9).
As a result the solution of the equations (2) has the 
following mul\-ti\-sca\-le decomposition via 
nonlinear high\--lo\-ca\-li\-zed eigenmodes
$$
F(t,x_1,x_2,\dots)=
\sum_{(i,j)\in Z^2}a_{ij}U^i\otimes V^j(t,x_1,\dots)
$$
\begin{eqnarray}
V^j(t)=
V_N^{j,slow}(t)+\sum_{l\geq N}V^j_l(\omega_lt), \ \omega_l\sim 2^l 
\end{eqnarray}
$$
U^i(x_s)=
U_M^{i,slow}(x_s)+\sum_{m\geq M}U^i_m(k^{s}_mx_s), \ k^{s}_m\sim 2^m
$$
%%\end{eqnarray}
%%}
which corresponds to the full multiresolution expansion in all underlying time/space 
scales.
The formulae (11) give the expansion into a slow 
and fast oscillating parts.  So, we may move
from the coarse scales of resolution to the 
finest ones for obtaining more detailed information about the dynamical process.
In this way one obtains contributions to the full solution
from each scale of resolution or each time/space scale or from each nonlinear eigenmode.
It should be noted that such representations 
give the best possible localization
properties in the corresponding (phase)space/time coordinates. 
Formulae (11) do not use perturbation
techniques or linearization procedures.
\begin{figure}[htb]                                                                    
\centering                                                                             
\includegraphics*[width=55mm]{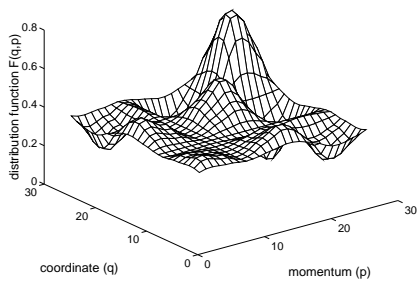}                                         
\caption{Localized mode contribution to distribution 
function.}                                                        
\end{figure}  
\begin{figure}[htb]
\centering
\includegraphics*[width=55mm]{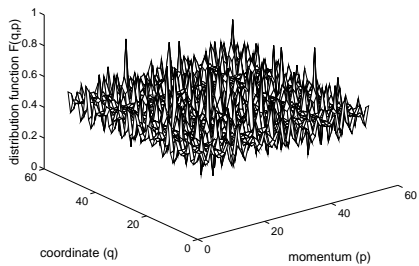}
\caption{Chaotic-like pattern.}
\end{figure} 
\begin{figure}[htb]
\centering
\includegraphics*[width=55mm]{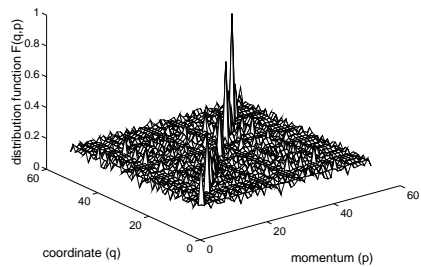}
\caption{Localized waveleton pattern.}
\end{figure} 
Numerical calculations are based on compactly supported
wavelets and related wavelet families [3] and on evaluation of the 
accuracy on 
the level $N$ of the corresponding cut-off of the full system (2) 
regarding norm (4):
$
\|F^{N+1}-F^{N}\|\leq\varepsilon.
$
We believe that the appearance of nontrivial localized patterns (a)-(c) demonstrated on
Fig.1--Fig.3 
constructed by these methods is a general effect which is also present in the 
full BBGKY hierarchy, due to its complicated 
intrinsic multiscale dynamics 
and it depends on neither the cut-off level nor the phenomenological-like
hypothesis on correlators. So, representations like (11) and the prediction of the existence of
the (asymptotically) stable localized patterns/states (energy confinement states) 
in BBGKY-like systems are the main results of this paper.
  
We are very grateful to Prof. Kaneko (KEK) and Prof. Perret-Gallix (CNRS) for kind help
and attention during ACAT03 Workshop at KEK.

\end{document}